\documentclass[a4paper]{article}
\usepackage{CJK}
\usepackage{amssymb}
\usepackage{graphicx}
\usepackage{booktabs}

\author{
Yichen Jiang\thanks{Y. Jiang and T. Liu are with the Department of Computer and Information Technology, Beijing Jiaotong University, Beijing, China, e-mail: (12112066@bjtu.edu.cn).}, Yi Ji\thanks{Y. Ji is with the Ira A. Fulton Schools of Engineering, Arizona State University, Arizona, USA.}, and Tianhua Liu$^*$
}

\title{An Anonymous Communication Scheme based on Ring Signature in VANETs}

\begin{document}

\maketitle

\begin{abstract}
Vehicular ad hoc networks allow vehicles to connect themselves as networks so that cars could communicate with each other. This paper introduces an anonymous communication scheme providing integrity protection, multi-level privacy and auditability. The scheme is based on a certificateless ring signature proposed in this paper, which is contributed to reduce the length of the signature and simplify the key management. In our scheme, vehicles can compose the anonymous group without the help of road-side infrastructure or central authority. The computation overhead is close to a normal signature scheme, so it is efficient in most application scenarios. We also present a small-scale implementation to show the availability of the prototype system.
\end{abstract}


\section{Introduction}
Intelligent transportation system promises to change our life in the future, and Vehicular Ad Hoc Networks (VANETs) are extremely important components of the intelligent transportation system. VANETs allow vehicles to connect themselves as networks in a self-organizing way so that vehicles can communicate freely with other vehicles or roadside infrastructures. People get better driving experience from VANETs.

There are two promising applications in vehicular communication system: vehicular warning and vehicular announcement \cite{DDSV09}. Both data integrity and auditability are of great importance in these applications. To achieve the goals of data integrity and auditability, digital signature methods can be utilized. However, naively using these mechanisms may affect the user's personal privacy. Cars are usually private possessions and kept by the owner for a long time, so one's car is likely to become a symbol of the owner. The data generated by the cars in VANETs, such as position information, traveling routes and even drive habits, are really sensitive for the owners.

We consider a real deployed system may involve more problems except for the requirements of data integrity, auditability and privacy. The first one is the real-time processing of the data \cite{DDSV09}\cite{HL08}. Data in some application scenarios must be processed very quickly, thus complicated cryptographic algorithms are not appropriate; The second problem is the adjustment of anonymity in the system. In some systems, not every user need to be anonymous, and the user may not keep anonymous or non-anonymous all the time \cite{RNL11}. For example, a police car on duty is required to make its identity public. However, when it is tracing a criminal secretly, it must conceal its identity and be anonymous as the cars of the public; The third one is the problem of central registration. Consider the following scenario: a car from Shanghai comes to Beijing. This car was registered in a different institution with cars in Beijing, so it is very easy to distinguish the messages coming from this car from the messages from Beijing's cars (maybe because it was issued a different root certificate with the cars in Beijing). This car must be re-registered in Beijing to keep anonymous, but it takes a lot to be done. In order to minimize the incidence of re-registration, as many cars as possible need to be registered in the same institution. However, it is not only resulted in management difficulties, but also not conducive to bootstrap the system \cite{PP05}.

Existing works cannot meet all the demands aforementioned. Pseudonym-based schemes usually have the advantage of efficiency, and a recent proposal of Bhavesh et al. achieved multi-level privacy by setting different lifetime of pseudonyms \cite{BMH13}. However, the scheme needs support of stable and secure roadside infrastructures. It is not only inconvenient to deploy the system but also takes a lot of burden of communication and computation to the roadside infrastructures. Group signature is another common approach used to implement vehicular anonymous communication. The drawback of group-signature-based schemes is the computation overhead. Although later works try to reduce the complexity of the operations \cite{ZPL12}\cite{ZX13}, they cannot satisfy the real-time processing of the vehicular warning application. Besides, limited by the features of group signature, group-signature-based schemes cannot address the problem of multi-level anonymity and central registration. To achieve multi-level privacy-preservation, Xiong et al. proposed a ring-signature-based scheme in \cite{XCL12}. We consider there is a serious security problem in Xiong's scheme because it is vulnerable to the unauthorized participation attack. An illegal user may compute key pairs by himself without the authentication of the Member Manager, and the other users cannot know if the public key is coming from a legitimate user.

In this paper, we proposed an anonymous communication scheme based on ring signature. We firstly designed a certificateless ring signature scheme based on the generalized ring signature proposed by Ren and Harn \cite{RH08}. The proposed ring signature scheme is contributed to reduce the length of the signature and simplify the key management since user IDs are used to compute a signature instead of a long string of public keys and corresponding certificates. Secondly, we present an anonymous vehicular communication scheme based on the proposed ring signature scheme. A protocol is designed in our scheme to achieve distinguishability of origins and non-repudiation. With hierarchical key structure, the average overhead of our scheme is close to a normal signature scheme. Finally, we implemented a small-scale prototype system on the PC platforms. The test results of the prototype system show the availability and the robustness of the proposed system.

Our solution has the following advantages: (1) It is very efficient and can satisfy the requirement of most application scenarios; (2) Our scheme provides multi-level privacy-preservation, so different users will have different anonymous degrees. The anonymity of the user is adjustable, and the user can decide his anonymity freely by dynamically setting up the ring; (3) It is decentralized. There is no registration center, and the vehicles compose the anonymous group in a self-organizing way.

The rest of this paper is organized as follows. In Section 2, we give a survey of prior works. The background of our research is given in Section 3. We then describe our scheme in detail in Section 4. In Section 5, we analyze the security and performance of our scheme, and present a comparison between our scheme and some of the existing works. Finally, we conclude this paper in Section 6.
\section{Related Work}
Many schemes have been proposed to provide data integrity and privacy protection in vehicular communication system. According to the approach employed in the scheme, they can be classified into three types: pseudonym-based scheme, group-signature-based scheme, and hybrid scheme.

Pseudonym-based schemes (\cite{BMH13},\cite{KWL09},\cite{PBHS08},\cite{RH07}) mainly rely on the public key cryptography. The so-called pseudonym, is virtually a short-term certificate which does not contain the identity information. The vehicles sign a message with its private key, and the others verify the signature of the message with the pseudonym. The generator of key pairs may be every single vehicle \cite{PBHS08}, or a public CA \cite{KWL09}.  Vehicles achieve the purpose of anonymity by constantly changing the key pairs. The issuer of pseudonyms is CA, so the CA in pseudonym-based schemes can reveal the real identity of the user to achieve auditability.

Pseudonym-based schemes are usually simple and efficient, and can be applied in a variety of scenarios in vehicular communication system. However, the main drawback of pseudonym-based schemes is the complicated management of pseudonyms. In \cite{RH07}, the CA needs to generate a large number of key pairs and pseudonyms, and preset them in the vehicles. The amount of key pairs must be enough until the next vehicle maintenance. Therefore, the computation burden of the CA and the memory requirement of the vehicle is very high in this way. \cite{KWL09} improved the scheme of \cite{RH07} by reducing the amount of certificates and computation burden, but part of the scheme is showed to be insecure \cite{CN10}. In \cite{PBHS08}, the vehicles are able to generate key pairs with the help of roadside infrastructures. The CA issues the certificate for the public key generated by car, but in this way, an all time online CA and a secure channel are required to keep the system works. Recently, Bhavesh et al. proposed a roadside-unit-aided scheme providing multiple levels of anonymity in \cite{BMH13}. The level of anonymity is determined by the number of pseudonyms and the lifetime of each pseudonym. However, the scheme needs support of stable and secure roadside infrastructures providing security services. It is not only inconvenient to deploy the system but also takes a lot of burden of communication and computation to the roadside infrastructures.

Group-signature-based schemes (\cite{CNW11},\cite{CMM14},\cite{LSHS07},\cite{MNL13},\cite{ZPL12},\cite{ZX13}) allow every legal user (vehicle) representing an organization (group) to sign a message. The group signature can be verified with the group certificate so that it is unclear that who the real signer is. The group manager is in charge of issuing the public parameters of the group, and every single user can join the group organization by running the join protocol. In general group signature schemes, the group manager is able to reveal the actual identity of the group members to achieve auditability.

Group-signature-based schemes require more complicated cryptographic operations, so the processing speed of these schemes is usually slow. The later proposals try to reduce the computational complexity of the group signature. \cite{ZX13}, \cite{CMM14}uses a simpler algorithm and batch verification technique to process large-scale verifications. In \cite{ZPL12}, powerful roadside units help vehicles to perform some complicated operations so as to improve the efficiency. However, the requirement of real-time data processing is not satisfied, for the delay of vehicular warning application is no more than 100 ms \cite{HL08}. Proposal \cite{CPHL07} presented a hybrid scheme based on the pseudonym and the group signature. In \cite{CPHL07}, the group signing key is used to generate a self-signing pseudonym certificate, and the pseudonym is used to communicate. This approach not only avoids the frequent computation of the group signature, but also solves the problem of pseudonym certificate generation. However, limited by the features of group signature, existing schemes cannot address the problem of anonymity adjustment and central registration.

A derived form of group-based scheme was proposed by Xiong et al., and they used a ring-signature-based scheme to achieve multi-level privacy-preservation \cite{XCL12}. The main disadvantage of Xiong's scheme is that the expensive operation of pairing used in the scheme is related to the ring size. If the ring size is big, the computation burden will be very high. Besides, there is a serious security problem of \cite{XCL12}. The key pair used in the ring signature is generated by the user himself, and then was sent to the Member Manager to verify. But an illegal user may compute key pairs by himself without the authentication of the Member Manager. The other users cannot know if the public key is coming from a legitimate user.

Moreover, a common issue of existing schemes is that of distinguishability of origin. The adversary may disguise himself as numerous vehicles. He can issue false messages, but the receiver cannot distinguish if these messages come from one car or many cars. For example, the adversary may pretend to be many cars and issue a message that there is a traffic jam, so the latter cars will be deceived to leave the road. Although proposal \cite{CNW11} applied the approach of one-time anonymous authentication to address the problem, we consider that there is still a question: in \cite{CNW11}, the two messages can be linked if and only if they are generated by the same user and the content of the messages are the same (for privacy consideration and solving the problem of disguising meanwhile), but the response of the vehicle is based on the meaning of the message. Therefore, the adversary can slightly modify the content of the message and keep the same semantics to deceive the threshold check.

We propose an efficient scheme that will not only achieve the objectives of security and privacy but also solve the problems of anonymity adjustment and central registration. The hierarchical design of master secret, self-signing certificates and pseudonym keys makes the average computation overhead of our scheme is close to a normal digital signature scheme. We proposed a communication protocol based on ring signature so that different users will have different anonymous degrees and the user can change his anonymity dynamically. Our scheme is immune from the typical attacks such as unauthorized participation, masquerade attack, Sybil attack in vehicular communication system. Central governmental authority does not participate in the deployment of our system, but just be an off-line supervisor. It matches the features of distributed and decentralized in VANETs.
\section{Background}
\subsection{Desired Requirements}
Before we have an introduction to our scheme, we firstly present the system goals of the scheme we designed. It includes two aspects: one is the basic security requirements in an anonymous communication system in VANETs, and the other is the unique features of our solution which can satisfy the special demands of some specific application scenarios.

Similar to most of the existing works \cite{LSHS07}\cite{PP05}\cite{XGH12}, the security requirements can be concluded as follows:
\begin{itemize}
\item \textbf{Integrity:} Includes data integrity and entity integrity. Data integrity ensures that the data can be transferred without unauthorized tampering. Entity integrity ensures the message comes from a legal source, and provides distinguishability of origin.
\item \textbf{Privacy:} Includes anonymity and unlinkability. The adversary cannot judge the sender of a message (anonymity), and cannot decide if any two messages come from the same source in case of user tracking (unlinkability).
\item \textbf{Auditablity:} Provides revocation and non-repudiation mechanism. If any entity is compromised, the other entities can identify the compromised entity and reject communication (revocation). The sender of a message cannot repudiate having sent the message to ensure accountable when an accident occurs.
\end{itemize}

Considering the requirements of the application scenarios, we expect to make our scheme satisfying the following features:
\begin{itemize}
  \item[-] \textbf{Efficient:} Messages can be dealt within a short time to meet the requirements of real-time processing.
  \item[-] \textbf{Anonymity adjustable:} The anonymity of the user is controlled by the user himself, and it depends on the actual situation.
  \item[-] \textbf{Decentralized:} Users compose the anonymous group in a self-organizing manner. There is no unified registration, and members of the group can be changed at any time.
  \end{itemize}
\subsection{Adversaries}
There are several possible attacks in a vehicular communication system, and we list some typical attacks here to help us analyze the security performance of our scheme. According to the power an adversary has, we classify the attacks in vehicular communication applications into two different types: attacks from the outside and attacks from the inside. Typical attacks from the outside include:
\begin{description}
  \item[Unauthorized participation] Unauthorized users directly or masquerade as a legitimate user to participate in the system, and affect the management of the system.
  \item[Message modification] The content or the source of the message is altered by malicious adversaries during the transmission.
  \item[Replay attack] Replay valid messages which were sent some time before. The adversary may avoid the authentication mechanism and bother the system, because these messages are coming from legitimate users.
  \item[Trace attack] By eavesdropping issued messages, the adversary tries to trace an entity. It infringes on the privacy of the user to bind the sensitive content of the messages and the identity of the user.
\end{description}

Attacks from the inside of the system mean that the adversary is more powerful. The adversary can even have a legitimate private key and certificates, but abuses the mechanism to attack the system. The masquerade attack and the Sybil attack are typical attacks from the inside.
\begin{description}
  \item[Masquerade attack] In some systems, roadside infrastructures (such as traffic lights) may provide convenience to some special vehicles (such as emergency vehicles), and the cars on the road will also make a way to these special vehicles voluntarily. Therefore, the adversary may masquerade as an emergency vehicle to obtain the priorities on road.
  \item[Sybil attack] The adversary may try to spread false messages to the others, and make the receivers believe that the messages come from different sources and the content of messages is true \cite{Dou02}. Because of the privacy preserving mechanisms, the receivers cannot distinguish the sources of the messages. Therefore, the adversary can mislead the other cars by making use of the contradiction between authentication and privacy.
\end{description}

\subsection{Technical Preliminaries}
The key technique we used in our scheme is the ring signature. The concept of ring signature was firstly proposed by Rivest, Shamir, and Tauman in 2001. It allows the actual signer to produce a signature with a group of public keys of other users (called non-signers) and his own private key. The cardinal part of the signature, the combining function, is to construct a ring structure for verification, so it is called ring signature. The verifier of the signature cannot tell who the actual signer is among the set of users in the signature so as to conceal the real identity of the actual signer. The most predominant advantage of the ring signature is that it can be produced in a self-organizing way. It can be used without registration, and the members and the size of the anonymous group can be decided by the actual signer himself. Operations of ring signature can be modeled as two algorithms:

The generalized ring signature scheme was proposed by Ren and Harn in 2008 \cite{RH08}. Different from the first ring signature scheme based on RSA, this scheme is based on the ElGamal signature scheme. The generalized ring signature simplifies the operation in Rivest's scheme, and all members can share the same domain by using the same prime modulus. Our scheme is proposed based on the generalized ring signature scheme, but is modified to a version supporting elliptic curves to get higher security level with shorter signature size. Besides, we simplify the management of public keys and shorten the length of the signature using the technique used in identity-based encryption (IBE), which is called Combined Public Keys. Our scheme will be presented in the next section.

The idea of IBE was proposed in \cite{Sha85}, 1984, and the first efficient scheme was proposed by Boneh and Franklin in 2001 \cite{BF01}. In an IBE system, a meaningful string can be used as the public key so that the user of the public key can easily learn the owner of a public key without a PKI certificate, because the public key itself may be the identity of its owner. The idea of IBE helps us to reduce the burden of certificates management, but the operations of most IBE schemes are relatively complicated. In \cite{LZ08}, Liu et al. present a simple IBE scheme using the technique of combined public keys. In this paper, we used the same notion as Liu el al. to improve the ring signature scheme.

\section{Proposed Scheme}
\subsection{Overview}
Different from existing works, our scheme is a hybrid scheme based on the generalized ring signature. There are some difficulties when applying the ring signature scheme in a real system. The first one is the problem of public key authentication. A basic solution in real systems is to use PKI certificates, but it brings extra computational cost. To address this problem, we proposed the certificateless ring signature scheme. It uses IBE techniques to turn the public key into a meaningful string, so we can use the identity of another user as his public key. It also helps to reduce the length of the signature, and the signer could generate a signature without mutual communication with the other user or the CA. The second difficulty is to provide anonymity and auditability at the same time. The generalized ring signature scheme provides a \texttt{convert} algorithm which allows the actual signer to convert the ring signature generated by himself to a normal digital signature so as to achieve auditability. However, the algorithm is proved to be incorrect because any user in the ring can claim the ring signature is generated by him \cite{WZS09}. The third problem is how to prevent the Sybil attack. In order to solve the second and the third problems, we designed protocols based on the proposed ring signature scheme. We bind the identity with the generation time and the content of the certificate without leaking the real identity so as to achieve authentication and privacy at the same time. In Section 4.2, we will introduce the certificateless ring signature algorithm in detail. Then, in Section 4.3, we will present the Anonymous Vehicular Communication System, which satisfies the privacy, distinguishability and auditability simultaneously.

Participants in our scheme can be classified into three types: the deployers, the supervisors, and the users. The deployers initialize the system and every user by generating parameters and preset credentials. The key update in our system is in the charge of the deployers during the vehicle maintenance process, which can be easily done by re-initializing the user. The supervisors can implement the auditability. They are able to issue key revocation list and trace vehicles for law enforcement purposes. The user of the system may be a normal private car, or a special vehicle on official duties.

A basic assumption of our scheme is that a vehicle is equipped with a security hardware module, which is a common method in vehicular communication system for security protection. The secure hardware modules can be treated as a black box. When giving a command, it will always give us a response correctly according to the algorithm. These secure hardware module has the capability to provide secure storage for private keys and perform cryptographic operations securely. In addition, the secure hardware module offers a trusted clock which cannot be compromised to the attacker. The credentials of the system are also stored in the hardware module, and the pseudonyms in the communication system will be cached so that it is not necessary to transfer the pseudonym credentials every times.

\subsection{Certificateless Ring Signature}
 We proposed a certificateless ring signature scheme based on the generalized ring signature scheme using the technique of Combined Public Key. The proposed ring signature scheme is contributed to reduce the length of the signature and simplify the key management. In our scheme, User IDs are used to compute a signature instead of a long string of public keys and the corresponding certificates. The scheme includes the \textbf{Setup}, \textbf{Key Generation}, \textbf{Ring Sign}, and the \textbf{Ring Verify} algorithms. The detailed description of the proposed ring signature scheme is given below.
 \\
\textbf{Setup}\indent Let $\mathbb{G}$ be an addition group consisting of points on an elliptic curve and the order of $\mathbb{G}$ is $q$. Let $P$ be a generator of $\mathbb{G}$. Firstly, select $n$ secret values $x_i \in \mathbb{Z}_{q}^{*}(1\leqslant i\leqslant n)$ randomly, and compute $Y_i = {x_i} \cdot P(1\leqslant i\leqslant n)$ for each $x_i$. Secondly, select two hash function $H_0: \{0,1\}^* \rightarrow \{0,1\}^n, H_1:\mathbb{G} \rightarrow \mathbb{Z}_{q}$. Define $X = (x_1,x_2,\cdots, x_n)$ as the master private key vector, and define $Y = (Y_1,Y_2,\cdots, Y_n)$ as the master public key vector. Finally, set the system public parameters to be $(\mathbb{G}, P, q, Y, H_0, H_1)$ and make them public. \\
\textbf{Key Generation}\indent For every user with identity $id$, the private key of the user is
\begin{displaymath}
    d_{id} = \sum_{i=1}^n h_ix_i \bmod q
\end{displaymath}
where $h_i$ is the $i$th bit of $H_0(id)$, $i = 1,\cdots, n$.\\
\textbf{Ring Sign}\indent Suppose $m$ is the message to be signed, and the identity of the signer is $id_{s}$. Thus, the private key of the signer is $d_{id_s}$. When he wants to generate a signature, the signer computes as follows:
\begin{enumerate}
  \item \textbf{Pick $r-1$ valid identities.} The signer picks another $r-1$ valid users with himself to form a ring of $r$ members. The method to pick identities will be explained in the next section, and here we concentrate on the algorithm itself. Note that $U = \{id_{1},id_{2},\cdots,id_{s},\cdots,id_{r}\}$ is the sequence of the identities of the ring members, and the identity of the signer is the $s$th identity in the sequence.
  \item \textbf{Extract public key.} For each $id_{i}$ in $U$, the signer computes
        \begin{displaymath}
        E_{id_{i}} = \sum_{j=1}^n h_{j} \cdot Y_{j}
        \end{displaymath}
        where $h_j$ is the $j$th bit of $H_0(id_i)$. The sequence of public keys of all the ring members is $\{E_{id_1},E_{id_2},\cdots,E_{id_s},\cdots,E_{id_r}\}.$ The calculated public keys can be cached so that it does not have to be recalculated every time.
  \item \textbf{Create forgeries.} For each identity $id_{i}$ in $U$, $id_i \neq id_s$, the signer picks $a_{id_i}, b_{id_i} \in \mathbb{Z}_{q}^{*}$ randomly and computes forgery with the public key $E_{id_i}$ as follows:
        \begin{displaymath}
        U_{id_i} = {a_{id_i}} \cdot P + {b_{id_i}} \cdot {E_{id_i}}
        \end{displaymath}
        \begin{displaymath}
        v_{id_i} = -H_1(U_{id_i}) (b_{id_i})^{-1} \bmod q
        \end{displaymath}
        \begin{displaymath}
        m_{id_i} = a_{id_i} v_{id_i} \bmod q
        \end{displaymath}
        It can be proved that ($U_{id_i},v_{id_i}$) is a valid ElGamal signature of message $m_{id_i}$ \cite{ElG85}, and the tuple $<m_{id_i},U_{id_i},v_{id_i}>$ is the forgery for member $id_i$.
  \item \textbf{Initialize the ring equation.} Pick a random value $\gamma \in \mathbb{Z}_{q}^{*}$ for the ring equation.
  \item \textbf{Solve the verification equation.} Note that $m$ is the message to be signed. Use a hash function $h$ which was described in \cite{BSS02} to construct the ring equation as follows:
        \begin{displaymath}
        w_{id_{s+1}} = h(m,\gamma)
        \end{displaymath}
        \begin{displaymath}
        w_{id_{s+2}} = h(m,w_{id_{s+1}} \oplus m_{id_{s+1}})
        \end{displaymath}
        \begin{displaymath}
        \vdots
        \end{displaymath}
        \begin{displaymath}
        w_{id_{s-1}} = w_{id_{s+r-1}} = h(m,w_{id_{s-2}} \oplus m_{id_{s-2}})
        \end{displaymath}
        \begin{displaymath}
        w_{id_s} = w_{id_{s+r}} = h(m,w_{id_{s-1}} \oplus m_{id_{s-1}})
        \end{displaymath}
        In order to glue the ring, which is $w_{id_s} \oplus m_{id_s} = \gamma$. Therefore,
        \begin{displaymath}
        m_{id_s} = \gamma \oplus w_{id_s}
        \end{displaymath}
  \item \textbf{Sign $m_{id_s}$ with signer's private key.} The actual signer generates the ElGamal signature of his $m_{id_s}$. To form a signature on specified value, the signer must use his private key $d_{id_s}$. Firstly, the signer picks $l$ uniformly from $\mathbb{Z}_q$. Then, he computes
        \begin{displaymath}
        U_{id_s} = l \cdot P
        \end{displaymath}
        \begin{displaymath}
        v_{id_s} = (m_{id_s} - d_{id_s} H_1(U_{id_s})) l^{-1} \bmod q
        \end{displaymath}
        Finally, ($U_{id_s},v_{id_s}$) is a valid ElGamal signature of message $m_{id_s}$, and output the tuple $<m_{id_s},U_{id_s},v_{id_s}>$ for the actual signer.
  \item \textbf{Output the ring signature.} The signature on the message $m$ is:\\
        \begin{center}
        $S = (x,w_{id_x};id_1,id_2,\cdots,id_r;<m_{id_1},U_{id_1},v_{id_1}>,<m_{id_2},U_{id_2},v_{id_2}>,\cdots,<m_{id_r},U_{id_r},v_{id_r}>)$\\
        \end{center}
        where $x$ is randomly selected among $1,2,\cdots,r$.
\end{enumerate}
\textbf{Ring Verify}\indent Given message $m$ and its corresponding signature $S$, the verifier can verify the validity of $S$ as follows:
\begin{enumerate}
\item \textbf{Extract public key.} Get the public key sequence of all ring members $\{E_{id_1},E_{id_2},\cdots,E_{id_s},\cdots,E_{id_r}\}$ with the same method as the extracting public key period in ring signing.
\item \textbf{Verify the signature of each tuple.} For each $id_i, i = 1,2,\cdots,r$, verify the ElGamal signature of every tuple with the public key $E_{id_i}$. The verifier checks the following equation:
    \begin{displaymath}
    {m_{id_i}} \cdot P = H_1({U_{id_i}}) \cdot E_{id_i} + {v_{id_i}} \cdot {U_{id_i}}
    \end{displaymath}
    If any one of the tuples $<m_{id},U_{id},v_{id}>$ does not satisfy the equation, the verifier rejects the signature $S$.
\item \textbf{Verify the ring equation.} The verifier checks that if all the $m_{id_i}$ together can form the ring equation. Using the start $x$ and its corresponding value $w_{id_x}$, the verifier checks the following equation:
    \begin{displaymath}
    w_{id_x} = h(m,m_{id_{x+r-1}} \oplus h(m,m_{id_{x+r-2}} 
    \end{displaymath}
     \begin{displaymath}
    \oplus h(m,\cdots \oplus h(m,m_{id_x} \oplus w_{id_x})\cdots)))
      \end{displaymath}
    where $h$ is the same hash function as the solving verification equation period in ring signing. If the ring equation is satisfied, the verifier accepts the signature $S$. Otherwise, reject $S$.
\end{enumerate}
\subsection{Anonymous Vehicular Communication System}
We introduce the proposed Anonymous Vehicular Communication System in this section, which is designed based on our certificateless ring signature scheme. To simplify the explanation, we use the inter-vehicles communication as an example, and it is similar for the situation of car-roadside-unit communication. We suppose that the deployers of the system are the independent car manufactories, and the government which does not participate in the deployment of the system is acting as the supervisor. There are seven periods corresponding to different operations in the system. Operation \textbf{Init} is used by the manufactories to initialize the system parameters, and operation \textbf{Join} is used to pre-distribute keys and credentials to every individual vehicle released by the manufactory. Operation \textbf{Reveal} is designed for the supervisor to identify the vehicle for law enforcement purposes.  The other operations, \textbf{genPseudonym}, \textbf{genMessage}, \textbf{Send}, and \textbf{Receive} are executed by the vehicles in the system. These algorithms are described in the following subsections respectively.
\subsubsection{Init}
This algorithm was run by each car manufactory to set the Anonymous Vehicular Communication System up. Firstly, all manufactories discuss the security level and the addition group $\mathbb{G}$ on an elliptic curve. Then, every manufactory runs the algorithm \texttt{Setup} in the proposed ring signature scheme according to the group $\mathbb{G}$ respectively. Denote $q$ as the order of group $\mathbb{G}$, and $P$ as the generator of $\mathbb{G}$. Thirdly, choose three hash functions $h_0, h_1: \{0,1\}^* \rightarrow \mathbb{G}, h_2: \{0,1\}^* \rightarrow \mathbb{Z}_{q}$.  In implementation, they can be the same hash function. Finally, make all generated parameters and selected functions public. After the decision of prime $p$, every manufactory can issue the parameters and run the \textbf{Join} algorithm introduced in the next part independently, so it is conducive to the deployment of the system.
\subsubsection{Join}
Once a car is produced, the manufactory uses this algorithm to initialize the vehicle hardware module so that the car is able to join the communication system. Every car generates a master secret $f \in \mathbb{Z}_q$ at the very beginning and saves it at the security hardware module. The master secret is unique,and it is protected by the hardware from leaking to the others. Then, the manufactory runs the \texttt{Key Generation} algorithm to compute the private key $d_{id}$ for the car, and preset it in the hardware module. A simple method is to use the license plate number as the input of the \texttt{Key Generation} algorithm. Finally, store the master public key vectors $Y$ of different manufactories in the secure hardware module.
\subsubsection{genPseudonym}
This algorithm is used to generate a pseudonym in the process of communication for the car. The pseudonym virtually is a public key certificate that does not contain the identity information, and usually has a short life cycle. Public keys can be authenticated by the pseudonym without the leak of the identity information. The algorithm of pseudonym generation in our system is described as follows:

\begin{center}
\begin{tabular}{l}
  \toprule
  \texttt{INPUT:} $f; U = \{id_1,id_2,\cdots,id_s,\cdots,id_r\}; d_{id_s}$.\\
  \midrule
  $pk,sk \leftarrow genKeyPair()$ \\
  $C \leftarrow pk||\cdots||expiration;$\\
  $H \leftarrow \lfloor \frac{currentTime}{minSpanTime} \rfloor$\\
  $J \leftarrow h_0(C); K \leftarrow h_1(H)$\\
  $R \leftarrow f \cdot J; T \leftarrow f \cdot K$ \\
  $L \leftarrow h_2(C||R||T)$ \\
  $S \leftarrow Ring\textrm{-}sign(L,U,d_{id_s})$\\
  \midrule
  $\texttt{OUTPUT:} \sigma \leftarrow <C,R,T,S>$\\
  \bottomrule
\end{tabular}
\end{center}

In the inputs of the algorithm, $f$ represents the master secret stored in the secure hardware module, and $d_{id_s}$ is the private key of the user $id_s$. $U$ represents the union of a series of user IDs with which the car $(id_s)$ together composes the anonymous group. The members in $U$ and the size of $U$ can be decided by the user at any time. Note that special vehicles on duties must make its identity public, which means to set the car itself as the only user in $U$. We suppose there is a buffer in which some user IDs are preset, and the IDs in $U$ are coming from this buffer. Every time the vehicle generates a pseudonym certificate, it will randomly select some IDs in the buffer. During the trip, the IDs in the buffer will be constantly updated by randomly picking some IDs in ring signatures sent by others and saving it into the buffer.

Firstly, the car generates a transient key pair used to identify the messages. The private key $sk$ of the transient key pairs is stored in the hardware module to use, and the hardware module issues the pseudonym certificate to the public key $pk$. Secondly, put the public key value, certificate expiration, etc. to construct the content of pseudonym $C$. Then, compute the additional time stamp $T$. $currentTime$ is the current system time, and $minSpanTime$ is the minimum generation interval of the pseudonym certificate. Therefore, the certificates generated at the same period will have the same value of $T$. Next, we bind the $C,T$ with the master secret of the secure hardware module. Finally, sign the generated values with the ring signature, and output the pseudonym certificate $\sigma$. The execution of the algorithm is finished by the secure hardware module.

\subsubsection{genMessage}
This algorithm is used to generate application messages. The output is $msg = <M,N>$, where $M$ is the content of the message, $N$ is the authentication part of the message. It is the signature of $M$ with the generated transient private key $sk$, or $N = Sign(M,sk)$ in mathematics. The execution of the algorithm is finished by the secure hardware module.
\subsubsection{Send}
Messages in our system are classified into two types: certificate messages and application messages. The generation algorithms have been given in last two sections. This algorithm is used to send a message. In theory, for every participant in the communication, the certificate message is only need to be computed and send once if the transient key pair remains unchanged. However, in order to enhance the robustness, we introduced the parameter $k$. It means the pseudonym certificate need to be re-sent after every consecutive $k$ application messages in case of incorrect receiving of the pseudonym certificate. The process of the \textbf{Send} can be described as follows:

\begin{center}
\begin{tabular}{l}
  \toprule
  \texttt{INPUT:} $\sigma; M = \{msg_1,msg_2,\cdots,msg_m\};k$.\\
  \midrule
  for $i = 1$ to $m$\\
  \{\\
  $\qquad$ if $i \% k = 0$ then\\
  $\qquad\quad$ \texttt{send}($\sigma$)\\
  $\qquad\quad$ \texttt{send}($msg_i$)\\
  $\qquad$ else\\
  $\qquad\quad$ \texttt{send}($msg_i$)\\
  \}\\
  \bottomrule
\end{tabular}
\end{center}

\subsubsection{Receive}
According to different message types, the vehicle responds distinctively. When a car receives a message coming from other cars, this algorithm is executed to deal with the packages.

\begin{center}
\scalebox{0.7}
{
\begin{tabular}{l}
  \toprule
  \texttt{INPUT:} $\sigma = <C,R,T,S>$ or $msg = <M,N>; $\\
  $\qquad\quad PseudonymBuf = \{{\sigma}_{1},{\sigma}_{2},\cdots\}; Rogue List = \{f_0,f_1,\cdots\}$.\\
  \midrule
  if($IsPseudonym()$)\\
  \{\\
  $\quad$if $\sigma = {\sigma}_{i}$ for any ${\sigma}_{i}$ in $PseudonymBuf$ then\\
  $\quad\quad$\texttt{return $\bot$}\\
  $\quad$if $T = T_i$ for any ${\sigma}_{i}$ in $PseudonymBuf$ then\\
  $\quad\quad$ \texttt{reject}\\
  $\quad J' \leftarrow h_0(C)$\\
  $\quad$if $R = {f_i} \cdot J'$ for any $f_i$ in $Rogue List$ then\\
  $\quad\quad$ \texttt{reject}\\
  $\quad$if $Ring\textrm{-}verify(h_2(C,R,T),S) = reject$ then\\
  $\quad\quad$ \texttt{reject}\\
  $\quad PseudonymBuf \leftarrow \sigma$\\
  $\quad$\texttt{accept}\\
  \}\\
  else if($IsMessage()$)\\
  \{\\
  $\quad {\sigma}_i \leftarrow PseudonymBuf$ for the corresponding pseudonym\\
  $\quad pk_i \leftarrow {\sigma}_i$\\
  $\quad$if $Verify(M,N,pk_i) = reject$ then\\
  $\quad\quad$ \texttt{reject}\\
  $\quad$\texttt{accept}\\
  \}\\
  else\\
  $\quad$\texttt{return $\bot$}\\
  \bottomrule
\end{tabular}
}
\end{center}

When the connection between two cars is established, the car usually receives the pseudonym certificate first to verify the application messages. Firstly, if the certificate has already been in the buffer, it omits the message. Secondly, by checking the additional information of $T$, the receiver could be sure if the certificates come from one source. If a certificate in the buffer has the same value of $T$, it means these two certificates may be coming from one car, and then it rejects the message. There may exist false alert in this approach. Because $f_1 = f_2 \Rightarrow f_1 \cdot T = f_2 \cdot T$, but $ f_1 \cdot T_1 = f_2 \cdot T_2 \nRightarrow f_1 = f_2$. The possibility of false alert is very low because of the randomness of hash functions and the master secret. In addition, if we consider the security of the clock time in the hardware module, then different cars will have the same value of $T$. When trying to make $T_1 \neq T_2$. The adversary must replay the message before. However, the replay attack can be easily prevented by checking the expiration of certificates. Thirdly, if the pseudonym comes from a compromised hardware module, the receiver can find out by the comparison of the revocation list. Finally, the validity of the certificate can be judged by the ring signature verification algorithm. After all these checks, the legitimate pseudonym will be appended into the buffer to verify application messages later. The ID series in $S$ may be saved in the ID buffer so that the receiving car could use these IDs to generate its pseudonym later.

As for application messages, the car firstly extracts the corresponding transient public key $pk$ from the certificates buffer. By verification algorithm,$N \stackrel{?}{=} Verify(M,pk)$, the integrity of the message can be verified. Then, it will be delivered to the application to respond.
\subsubsection{Reveal}
This algorithm is used to determine if the message comes from the suspected user. It can only be executed by the supervisor at a certain time, and it implements non-repudiation. When special accident occurs, the sender of a message may need to be identified for law enforcement purposes. Then, the supervisor has the right to use this operation of the security hardware module. For security concerning, the security hardware module will authenticate the identity of the sender before the \textbf{Reveal} operation is executed. Then, the security hardware module executes as follows: suppose the certificate to be identified is $\sigma = <C,R,T,S>$. Firstly, the supervisor initiates the \textbf{Reveal} request, and sends $C$ to the hardware module. After receiving the content, the security hardware module generates a certificate denoted as $\sigma' = <C,R',T',S'>$ for $C$. Then, the certificate is sent back. The supervisor verifies $R \stackrel{?}{=} R'$, and it proves if the certificate to be identified is coming from the participant car. Since $R = f \cdot h_0(C)$,if the values of $R,C$ are equal, the value of $f$ must be the same. The master secret identifies the only car. If the two certificates come from the same car, the car verified cannot deny having sent the message.

\section{Evaluation}
\subsection{Security Analysis}
In this section, we analyze that how the proposed scheme prevents the typical attacks described in Section 3.2 and satisfies the desired requirements. In our scheme, every message is attached with a digital signature, and can be verified with a corresponding certificate sent together with the message. This mechanism brings the features of integrity and authentication, and prevents the participation of unauthorized users and message modification. The replay attack can be avoided by checking the expiration of the certificate and the time information in the application packages since the security hardware module provides us with a secure time base.

In order to preserve the privacy of the users, these certificates in our scheme are authorized by the technique of ring signature. The members in the ring are randomly selected and constantly updated. The adversary can only be sure that a certificate comes from one of the members in the ring so that achieve the goal of anonymization. The messages in our scheme are just able to be linked in a pseudonym period. By changing the pseudonym certificate from time to time, it protects the user from being traced. The ring signature also contributes to preventing the masquerade attack because of anonymity-adjustment. In our scheme, special vehicles must make its identities public to gain roads privilege. It means that the special vehicle is the only member of the ring. Therefore, an adversary without a real private key of a special vehicle cannot generate a valid ring signature for the certificate.

The situations of non-repudiation and distinguishability of origin are more complicated. We introduce two extended parts into the pseudonym certificate to support the function. One ($R$) is used to bind the content of the certificate with the identity of the signer, and the other ($T$) is used to bind the issue time of the certificate with the identity. Note that the identity of the signer is represented by a secret stored in the security hardware module and the two parts of the certificate are present in the form of discrete logarithm. According to the decisional Diffie-Hellman (DDH) assumption, it is hard to tell ${xy} \cdot P$ from a random string $z \cdot P$ for any polynomial algorithm even if $x \cdot P$ and $y \cdot P$ are given. Therefore, if the adversary wants to trace the user from the information included in the extended parts, he is able to solve the DDH problem. Based on the extended part $R$, we could implement the \textbf{Reveal} algorithm as well as the revocation check in \textbf{Receive}. As for the Sybil attack, it was avoided by the check of the extended part $T$ for the adversary abusing the algorithm \textbf{genPseudonym} to generate many certificates and pretend to be a number of identities in a very short time (a $minSpanTime$ period) can easily be detected by comparing the the extended part $T$.

We compare our scheme with two existing schemes. One is the TAA scheme \cite{CNW11} for being later and more completed, and the other is the RSB scheme which is similar with our scheme to use ring signature to achieve multi-level privacy \cite{XCL12}. The comparison result is given in Table 1. Data integrity and privacy are achieved in all the schemes, but the scheme in \cite{XCL12} cannot provide the entity integrity because it cannot prevent the Sybil attack. Auditability is also satisfied in TAA and our scheme while RSB is not. The additional application requirements of anonymity adjustment and multi-level privacy are not implemented in TAA, and the requirement of decentralization is only implemented in our scheme.

\begin{table*}
\caption{Comparison of the security goals.}
\begin{center}
\begin{small}
\begin{tabular}{c|c c c c c c c}
  \toprule
                         &    Data   & Dist.          &  Privacy  &      Audit-    &   Anonymity    &    Decentra-   \\
                         & Integrity & of origin      &           &     ability    &   Adjustment   &    lization    \\
  \midrule
  TAA \cite{CNW11}       & ${\surd}$ &   ${\surd}$    & ${\surd}$ &  ${\surd}$     & ${\texttimes}$ & ${\texttimes}$ \\
  RSB \cite{XCL12}       & ${\surd}$ & ${\texttimes}$ & ${\surd}$ & ${\texttimes}$ &    ${\surd}$   & ${\texttimes}$ \\
  Our Scheme             & ${\surd}$ &   ${\surd}$    & ${\surd}$ &  ${\surd}$     &    ${\surd}$   &   ${\surd}$    \\
  \bottomrule
\end{tabular}
\end{small}
\end{center}
\end{table*}

\subsection{Performance Analysis}
In order to show the availability and the efficiency, we implement the proposed Anonymous Vehicular Communication System prototype based on the third-party open source software PolarSSL version 1.3.2 \cite{POLARSSL}. The test platform is HP Compaq 8200 Elite SFF PC platform, which is equipped with Intel Core i5-2400 quad-core CPU, clocked at 3.10 GHz, 4 GB RAM. All the computers are connected with 802.11 USB wireless Ethernet adapters in ad hoc mode. We consider the computation resource is not the limitation of the VANETs, and it is not a dream that the computing devices equipped on vehicles are as powerful as a PC.

The most expensive operation in our scheme is the certificateless ring signature signing and verifying, and may become the bottleneck of the system. Therefore, we test the computation time of the ring signature algorithms in different security levels and different ring sizes to show the efficiency. It is unnecessary to use a very high degree of anonymity (large ring size) in normal application because the members of the anonymous group will be changed constantly. Every time the vehicle updates a new pseudonym, it will associate some other cars which can be in anywhere with the information it generated. After some rounds updating, there will be a lots of cars related to the information, and it is enough to hide the true identity of the real sender. A larger ring size brings higher increment of members in anonymous group, but brings heavier computation burden. Balancing requirements of anonymity and performance, we set the upper limit of the ring size in our test to be ten. The test results are given in Figure 1. We can learn that the computation overheads increase linearly with the increment of ring size. We use the curves NIST recommended, and they are widely used in real systems \cite{ANSI9.62}. Figures 1 shows us the computation overhead is very small, which shows the high efficiency of our proposed scheme. The operations of ring signature are only used to generate the pseudonym certificate, and only use once during a pseudonym period. Thus, the time consuming will be better from the view of the whole system.

\begin{figure*}
  \centering
  \includegraphics[width=0.8\textwidth]{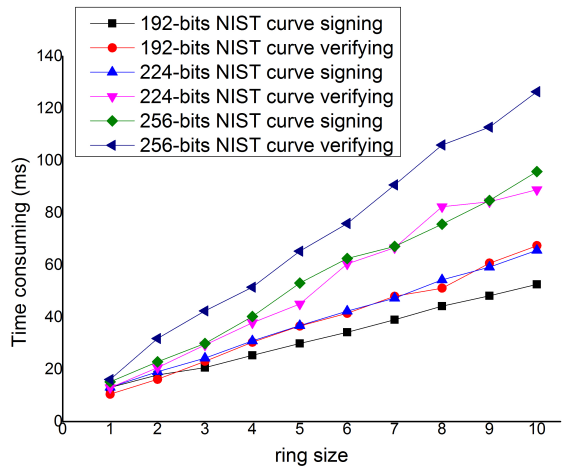}\\
  \caption{Time consuming of ring signing and ring verifying}
\end{figure*}

We then compare our proposed certificateless ring signature signing and verifying with the signing algorithm and the verifying algorithm in TAA scheme. Since the overheads of hash operation and other basic operations are very small, we estimate the computation overheads of the algorithms only depending on these complex operations such as scalar multiplication and pairing in elliptic curve. Consider a $r$ sized ring in our scheme. The signing operation needs $3r-1$ scalar multiplication operations, and the verifying operation needs $3r$ scalar multiplication operations. The TAA signing needs 6 scalar multiplications and 1 pairing, and the TAA verifying needs 5 scalar multiplications and 5 pairings. In usual implementation, the computation time of pairing is more than scalar multiplication. We test the computation time of scalar multiplication and pairing of library PBC \cite{PBC} implementation on type-D curves which are used in TAA scheme. We also test the computation time of scalar multiplication of library PolarSSL on NIST recommended curves which is used in our scheme at the same security level. According to our test of library PBC 0.5.14, the computation time of scalar multiplication is 1.92ms and the computation time of pairing is 12.93ms. The computation time of scalar multiplication on NIST curves is 2.05ms. Comparing the results, we can learn that our scheme has similar efficiency when the ring size is not too large. Note that the certificateless ring signature algorithm and verifying algorithm are computed only once for each pseudonym in the certificate message, and other application messages are signed by normal digital signature algorithm. If more application messages were sent, our scheme would reflect the greater advantage in efficiency.

Another comparison reference is the ring-signature-based scheme in \cite{XCL12} which is denoted as RSB. Similar ideas make the schemes more comparable, because the overhead of both of our scheme and RSB are proportional to the ring size. The signing operation needs $3r+3$ scalar multiplication operations, and the verifying operation needs $2r$ scalar multiplications and $2r+3$ pairings. RSB uses special curves (type-A1 curves) on which group elements have a order of a composite number. According to our tests on PBC, the computation time of scalar multiplication on type-A1 curves is 78.3ms, and the computation time of pairing is 152.1ms. Therefore, our scheme obviously has less computational complexity and is more efficient. Table 2 concludes the number of the expensive operations, where $\mathbb{P}$ denotes pairing, and $\mathbb{M}$ denotes scalar multiplication. Note that although we use the same symbol represent the operation of scalar multiplication, the computation time is different because these operations are running on different type of curves.

\begin{table*}
\caption{Number of expensive operations in different schemes.}
\begin{center}
\begin{small}
\begin{tabular}{c||c|c}
  \toprule
             &  Signing  & Verifying \\
  \midrule
  TAA        &$6\mathbb{M}+1\mathbb{P}$&$5\mathbb{M}+5\mathbb{P}$   \\
  RSB        &$(3r+3)\mathbb{M}$&$(2r+2)\mathbb{M}+(2r+3)\mathbb{P}$   \\
  Our Scheme &$(3r-1)\mathbb{M}$&$(3r)\mathbb{M}$                    \\
  \bottomrule
\end{tabular}
\end{small}
\end{center}
\end{table*}

The main disadvantage of our scheme is the length of the ring signature. A ten sized ring will generate a signature about 1KB. However, the transmission average speed of the wireless ad hoc network can achieve 740.07 Kb/s and even more according to our test. Therefore, it will be a very little delay and still acceptable comparing the TAA scheme at the aspect of communication transmission.

Table 3 shows the comprehensive data of the different operations. The data are measured under the 192-bits curves and the ring size $r = 10$. We set the parameter $k=10$, which means that every ten application messages are sent the pseudonym certificate needs to be resent once. We finally calculate the average cost of $n = 100$ messages according to the following formula and compare the average cost of the three schemes:
\begin{displaymath}
    \tau = \frac{1}{n} \times ({n \times t_{gM}} + t_{gP} + {n \times t_{sM}} + {\frac{n}{k} \times t_{sP}} + {n \times t_{vM}} + t_{vP}),
\end{displaymath}
where $t$ means the time cost. Footnotes $g,s,v$ represent for generating, sending and verifying, and footnotes $M,P$ represent for application Message and Pseudonym certificate. For example, the $t_{gP}$ means the time cost of generating a pseudonym certificate, and so on. The data of TAA and RSB we used in Table 3 are tested with library PBC 0.5.14 on our platform. Both PBC and PolarSSL are widely used open source library, so the test results are persuasive.

\begin{table*}
\caption{Comparison of time cost (ms) in different phases.}
\begin{center}
\begin{small}
\begin{tabular}{c||c|c|c|c|c|c||c}
  \toprule
             &  gen  & gen   & send & send & verify & verify & Average     \\
             &  Msg. & Pseu. & Msg. & Pseu.& Msg.   & Pseu.  & Cost($\tau$)\\
  \midrule
  TAA        &  24.4  &  N/A  &  1.9 & N/A  &  74.3  &  N/A   &   100.6      \\
  RSB        &  783   &  N/A  & 16.3 & N/A  &  1521  &  N/A   &   2320.3    \\
  Our Scheme &  2.1  & 52.6   & $<$1 & 9.7 &  6.7   &  67.4  &    10.9     \\
  \bottomrule
\end{tabular}
\end{small}
\end{center}
\end{table*}

From Table 3, we can see that the average computation cost of our scheme is the least, because most of the messages are processed by normal signature algorithm which is pretty fast and the ring signature is used only once. RSB is not very efficient because it needs the curve order to be a composite number. It takes large computation burden and size of elements when implementing the scheme, although it is provably secure without Random Oracle model. From the comparisons above, we can conclude that our scheme is very efficient and can satisfy the demands of different applications.

\section{Conclusion}
In this paper, we have presented a novel Anonymous Vehicular Communication Scheme based on ring signature. In order to meet the demands of some special application scenarios, we improved the generalized ring signature scheme with the idea of Combined Public Keys and used the proposed certificateless ring signature scheme in our system. Our scheme can satisfy the goals of authentication, multi-level privacy and auditability at the same time. Different from the existing scheme based on pseudonym or group signature, vehicles in our scheme can have different anonymity and can change its anonymity at any time. Besides, all the vehicles can compose an anonymous group in a self-organizing way without the help of a central authority. The analysis of security shows that our scheme is able to prevent the typical attacks in vehicular communication systems, and the efficiency of our scheme is also discussed. The computation overhead of our scheme is close to a normal signature scheme, so it is efficient in most application scenarios.

For future work, we intend to formally prove the security of the proposed scheme. Since giving the formal definitions of the diverse security requirements in VANETs is difficult, giving rigorous proofs of the proposed scheme is facing great challenges. Furthermore, some improvements for key revocation and malicious user tracing may also be worthy of studying.

\end{document}